\documentclass[a4paper,final]{appolb}
\pdfoutput=1
\usepackage[pdftex]{graphicx}
\usepackage[unicode=true, bookmarks=false,bookmarksnumbered=true,bookmarksopen=true, breaklinks=false,backref=false,pagebackref=false, colorlinks=true]{hyperref}
\hypersetup{pdftitle={ },pdfauthor={ },linkcolor=black,citecolor=black, urlcolor=black, filecolor=black,pdfpagelayout=OneColumn,pdfnewwindow=true,pdfstartview=XYZ, plainpages=false}
\usepackage{color}
\usepackage{array}
\usepackage{tabularx}
\usepackage{booktabs}
\usepackage{enumerate}
\usepackage{verbatim}
\usepackage{bm}
\usepackage{amsmath}
\usepackage{amssymb}
\usepackage{algorithm,algorithmic}
\usepackage{pgf}
\usepackage{verbatim}
\usepackage{braket}
\usepackage[T1]{fontenc}
\usepackage[english]{babel}
\usepackage{subfig}
\usepackage{float}
\usepackage{url}

\newcolumntype{Y}{>{\raggedright\arraybackslash}X}
\newcolumntype{W}{>{\raggedleft\arraybackslash}X}
\newcolumntype{Z}{>{\centering\arraybackslash}X}

\usepackage{verbatim}
\usepackage{multirow}

\begin{document}

\title{Study of compact U(1) flux tubes in 3+1 dimensions in lattice gauge theory using GPU's
\thanks{Presented by A. Amado at the International Meeting "Excited QCD", Peniche, Portugal, 6 - 12 May, 2012}}
\author{Andr\'{e} Amado, Nuno Cardoso, Marco Cardoso, Pedro Bicudo
\address{CFTP, Departamento de F\'{i}sica, Instituto Superior T\'{e}cnico, Universidade T\'{e}cnica de Lisboa, Av. Rovisco Pais, 1049-001 Lisbon, Portugal\newline}}

\maketitle

\begin{abstract}
We utilize Polyakov loop correlations to study (3+1)D compact $U(1)$ flux tubes and the static electron-positron potential in lattice gauge theory. By using field operators it is possible in U(1) lattice gauge theory to probe directly the electric and magnetic fields.
In order to improve the signal-to-noise ratio in the confinement phase, we apply the L\"uscher-Weiss multilevel algorithm.
Our code is written in CUDA, and we run it in NVIDIA FERMI generation GPU's, in order to achieve the necessary performance for our computations.
\end{abstract}
\PACS{11.15.Ha; 12.38.Gc}

\section{Introduction}

There are several models trying to describe confinement features that can achieve good successes, mainly dual superconducting models and effective string models. Though, the exact details of confinement mechanisms are still an open question in physics.

It is important to compare the predictions of these models with more fundamental calculations in order to check the domain of validity of the models and to understand more deeply the processes underlying confinement.

Studies have been conducted to pursue this objective with different gauge groups, mainly studying the potential of two static quarks. More recently, some studies have also focused the predictions for the tube flux itself. Usually this can only be done through the measurement of the squared field average, but in $U(1)$, being abelian, it is possible to study the electric and magnetic fields directly also.

\subsection{Effective string model}

According to the effective string model the quarks are bounded by a string that confines them in a asymptotically linear potential.

This model achieved several successes, being the most well-known the existence of the universal $1/r$ term in potential (L\"{u}scher term \cite{Luscher:1980ac}) which was confirmed in many studies regarding different gauge groups.

More precisely effective string model predicts the following potential at leading order \cite{Luscher:1980ac}
\begin{align}
 V(r) = \mu + \sigma r - \frac{\pi (d-2)}{24 r} + O(\frac{1}{r^3})
\end{align}
where $\mu$ is an arbitrary constant, $\sigma$ is the string tension and $d$ is the dimension of space.

It is now possible to check other predictions of this model including the ones related to the shape of the flux tube itself.

This have been done in 2+1 dimensions with several groups, since in 2+1 dimensions the theoretical predictions are well established, and the results have been found in agreement with this model. In 2+1 dimensions there are two results, for the limits of zero temperature and high temperature (although still in confining phase). These results consist of a logarithmic increase in the flux tube width at zero temperature and a linear increase at finite temperature (cf. \cite{Gliozzi:2010jh}), where the flux tube is evaluated according to the following definition
\begin{align}
w^2(r/2) = \frac{\int dx_\perp x_\perp^2 O(x_\perp)}{\int dx_\perp x_\perp^2}
\end{align}
where $O(x_\perp)$ is the profile of the field in the mediator plane of the charges.

\section{Methods}
We consider an euclidean 3+1 dimensional $U(1)$ lattice gauge theory with periodic boundary conditions. The action used is Wilson action for $U(1)$ 
\begin{align}
\beta S = \beta \sum_x \sum_{\mu < \nu} \left[ 1 - \mathcal{R}e \ U_{\mu\nu}(x) \right]
\end{align}
where $\beta=\frac{1}{g^2}$, $U_{\mu\nu}(x) = U_\mu(x)\ U_\nu(x+\hat{\mu})\ U^*_\mu(x+\hat{\nu})\ U^*_\nu(x)$ is the plaquette along the directions $\mu$ and $\nu$ and $U_\mu(x)$ is the gauge group element in position $x$ and direction $\mu$.

Wilson action is local (in the sense it only depends on plaquettes around it) which is a condition for using the multilevel method \cite{Luscher:2001up} we implement.

\subsection{Configurations}
Our lattice is an isotropic lattice of size $24^4$ and we use $\beta = 1$. This choice guarantees that we are working in confining region, at zero temperature, although with a $\beta$ big enough to have relatively stable results.

We use a combination of standard Metropolis and Overrelaxation algorithms to generate the configurations. As the time corresponding to the iterates between configurations is negligible compared to the time of needed for multilevel method we undertake more than enough iterates (100 iterates) in between any two used configurations to ensure those configurations are completely uncorrelated.

\subsection{Multilevel algorithm}
Multilevel algorithm was introduced by L\"{u}scher and Weisz \cite{Luscher:2001up} as a way to enhance the signal-to-noise ratio in large loop sizes. It consists in splitting the lattice in several layers where spatial links are fixed and performing partial updates of the rest of the lattice. 

This technique is very useful to study U(1) as the signal-to-noise ratio decreases extremely fast for increasing loop sizes in the confining region of this theory. Therefore we implement a two level multilevel algorithm \cite{Luscher:2001up,Koma:2003gi}, allowing us to calculate correlations at a distance up to 6 lattice spacings.

To calculate the potential we also replace the temporal direction links with the ones obtained from the analytical version of the multihit method \cite{Koma:2003gi,Parisi:1983hm}, which exhibit a smaller variance.

With $\langle P^*(0) P( r ) \rangle$ we can calculate the static potential
\begin{align}
V(r) = -\frac{1}{N_t} \ln \left[ \langle P^*(0) P( r ) \rangle \right] .
\end{align}

For calculating the electric and magnetic field we need to introduce a new operator. Electromagnetic tensor components can be probed with the following
\begin{align}
a^2 F_{\mu\nu} = \sqrt{\beta} \ \mathcal{I}m \ P_{\mu\nu} + \mathcal{O}(a^3)
\end{align}
and the squared components
\begin{align}
a^4 F_{\mu\nu}^2 = \beta \left[ 1 - \mathcal{R}e \ P_{\mu\nu} \right] + \mathcal{O}(a^6).
\end{align}

As we want the field in the presence of polyakov loop sources we should introduce them in the average
\begin{align}
\langle O(x) \rangle_{P^*P} = \frac{\langle P^*(0) P(r)\ O(x) \rangle}{\langle P(0)^* P(r) \rangle}-\langle O(x) \rangle
\end{align}
where $O(x)$ stands for any operator we want to measure.

To compute $\langle P^*(0) P(r)\ O(x) \rangle$ in the multilevel scheme we need to define a new operator that accounts for $O(x)$
\begin{align}
\mathbb{O}(x_0,x,t,r) = U^*_0(x_0,t)U_0(x_0 + r,t) O(x,t)
\end{align}
and 
\begin{align*}
\mathbb{TO}^{(2)}(x_0,x,t) = [\mathbb{T}(x_0,t,r) \mathbb{O}(x_0,x,t+1,r)] + [\mathbb{O}(x_0,x,t,r) \mathbb{T}(x_0,t+1,r)] .
\end{align*}
So we have for the average field (cf. figure \ref{fig:multilevel_field})
\begin{flalign}
\langle P^*(0)&P(r) O(x) \rangle = \frac{1}{V} \sum_{x_0}\sum_{t} \{ \hspace{7.5cm}\nonumber \\
\langle &[\mathbb{TO}^{(2)}(x_0,x,0)\mathbb{T}^{(2)}(x_0,2)] ... [\mathbb{T}^{(2)}(x_0,N_t-4)\mathbb{T}^{(2)}(x_0,N_t-2)] +... \nonumber\\
... + &[\mathbb{T}^{(2)}(x_0,0)\mathbb{T}^{(2)}(x_0,2)] ...[\mathbb{T}^{(2)}(x_0,N_t-4)\mathbb{TO}^{(2)}(x_0,x,N_t-2)]\rangle \}\nonumber
\end{flalign}
where $V$ is the total number of points in the lattice.

\begin{center}
	\includegraphics[width=0.9\textwidth]{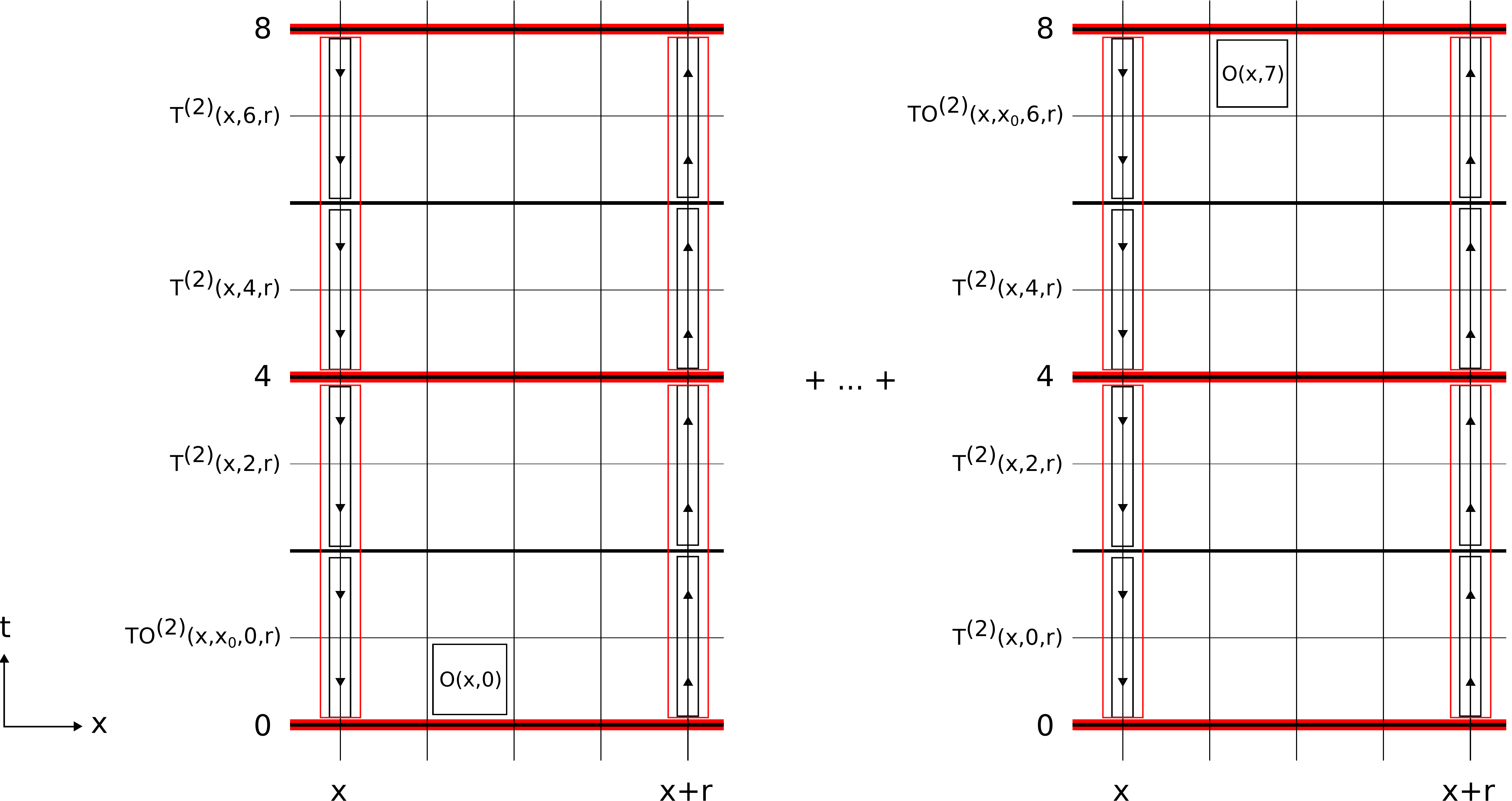}
	\captionof{figure}{Multilevel scheme for the field.}
	\label{fig:multilevel_field}
\end{center}

\section{Results}

\subsection{Potential}
We calculate the potential from 100 multilevel configurations, each one with 100 level 1 and 1000 level 2 multilevel iterates, with multihit method for further error reduction. We fit the results (figure \ref{fig:potential_graph}) and extract a value for string tension of $\sigma = 0.16719 \pm 0.00030$, if we force the L\"uscher term to be constant ($\chi^2/dof = 0.140$). With L\"uscher term as a fit parameter we obtain $\sigma = 0.1666 \pm 0.0022$ and L\"uscher term $0.2743 \pm 0.0605$ ($\chi^2/dof = 0.147$), in a good agreement with the expected value of $\pi/12 = 0.2618$. We do not include the first two points in the fit.

\begin{center}
	\includegraphics[width=0.75\textwidth]{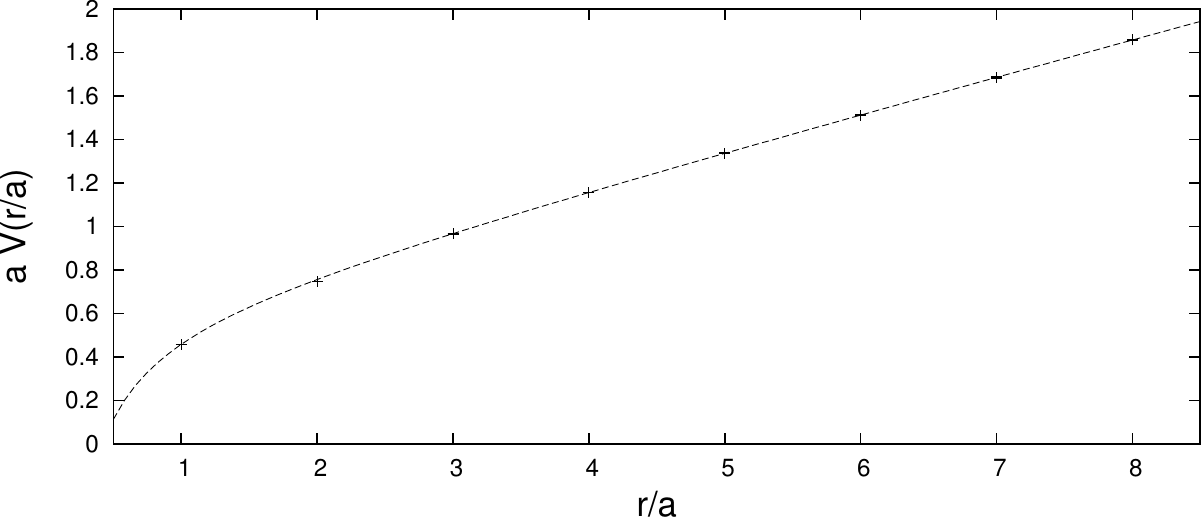}
	\captionof{figure}{Potential}
	\label{fig:potential_graph}
\end{center}

\subsection{Flux tube profile}
We calculate the $E_x$ tube profile in the middle plan between the charges. We use 100 multilevel configurations. The function fitted is the ansatz suggested in \cite{Gliozzi:2010jh}:
\begin{align}
\frac{\langle P^* P\ P_{\mu\nu} \rangle}{\langle P^* P \rangle} = A \exp (-x_\perp^2/s)\frac{1+B\exp (-x_\perp^2/s)}{1+D\exp (-x_\perp^2/s)}.
\end{align}
As expected we can notice that the flux tube (figure \ref{fig:flux_tube_profile}) gets broader at bigger charge separations. We quantify this result through the calculation of the flux tube width.
\begin{center}
	\includegraphics[width=0.75\textwidth]{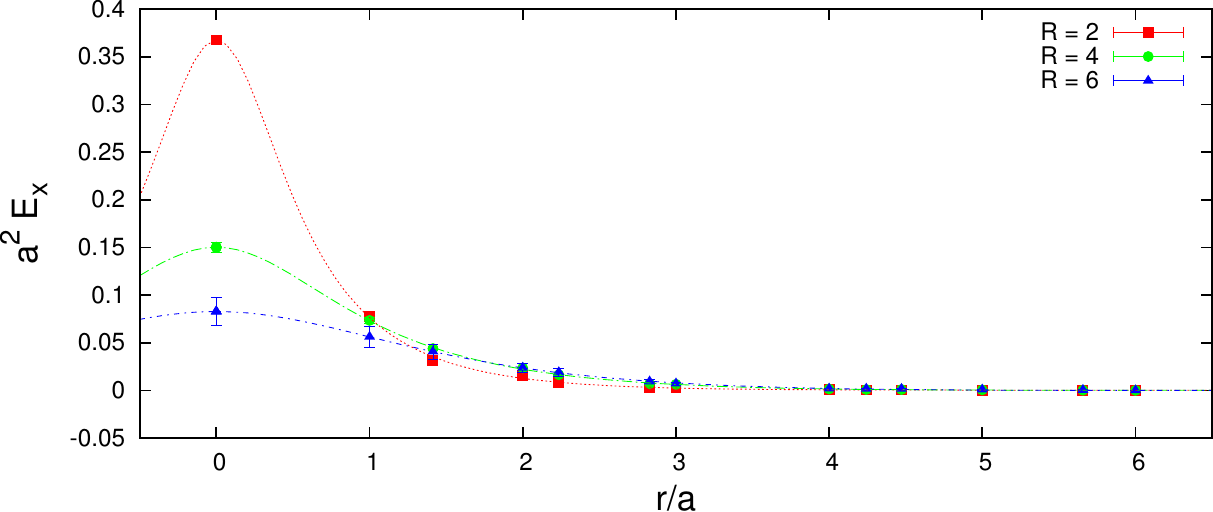}
	\captionof{figure}{Flux tube profile}
	\label{fig:flux_tube_profile}
\end{center}

\subsection{Flux tube width}
We integrate the flux tube ansatz fits to calculate the width (cf. table \ref{tab:electric_field_width}). The errors are estimated using a jackknife algorithm. As expected the flux tube width grows, although it is not possible to tell its exact form.

\begin{center}
	\begin{tabular}{l|c|c|}
		\multicolumn{1}{c}{
			\multirow{6}{*}{ \includegraphics[width=0.50\textwidth]{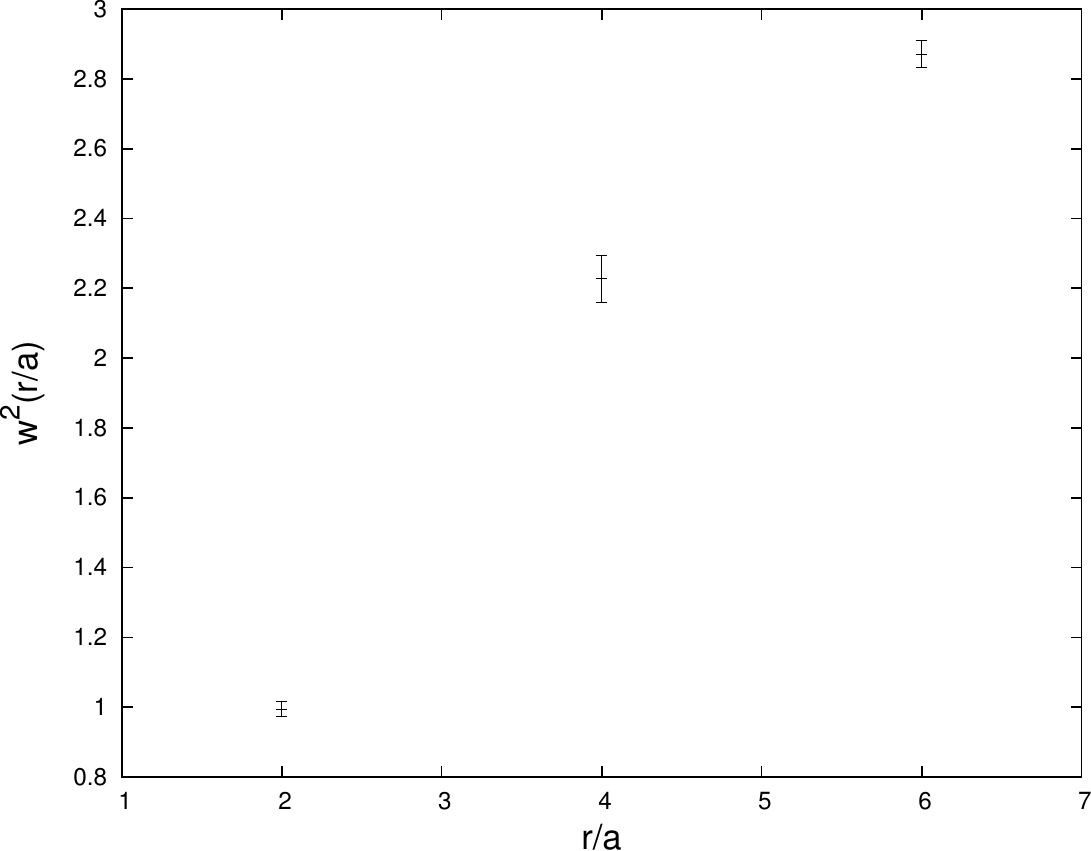} }} & \multicolumn{2}{c}{} \\
		\multicolumn{3}{c}{} \\
		\multicolumn{3}{c}{} \\
		\cline{2-3}
		\ &	r	& $w^2(r/2)$	\\ \cline{2-3}
		\ &	2	& $0.995 \pm 0.021$	\\ \cline{2-3} 
		\ &	4	& $2.227 \pm 0.067$	\\ \cline{2-3}
		\ &	6	& $2.871 \pm 0.039$	\\ \cline{2-3}
		\multicolumn{3}{c}{} \\
		\multicolumn{3}{c}{} \\
		\multicolumn{3}{c}{} \\
		\multicolumn{3}{c}{} \\
	\end{tabular}
	\captionof{table}{Electric field width}
	\label{tab:electric_field_width}
\end{center}

\section{Conclusions}
Our results for potential are in a good agreement with the predictions of the effective string model.

For flux tube profile we find an increasing width, but we are not able to tell the exact form of the broadening. Further research is needed, to increase the precision of the result with more configurations and the distance up to which we can compute the flux tube.

\section*{Acknowledgments}
This work was financed by the FCT contracts POCI/FP/81933/2007, CERN/FP/83582/2008, PTDC/FIS/100968/2008, CERN/FP/109327/2009, CERN/FP/116383/2010 and CERN/FP/123612/2011.
Nuno Cardoso is also supported by FCT under the contract SFRH/BD/44416/2008. Also we thank NVIDIA Academic Partnership for the support provided.

\bibliographystyle{elsarticle-num}
\bibliography{bib}

\begin{thebibliography}{1}
\expandafter\ifx\csname url\endcsname\relax
  \def\url#1{\texttt{#1}}\fi
\expandafter\ifx\csname urlprefix\endcsname\relax\def\urlprefix{URL }\fi
\expandafter\ifx\csname href\endcsname\relax
  \def\href#1#2{#2} \def\path#1{#1}\fi

\bibitem{Luscher:1980ac}
M.~Luscher, {Symmetry Breaking Aspects of the Roughening Transition in Gauge
  Theories}, Nucl.Phys. B180 (1981) 317.
\newblock \href {http://dx.doi.org/10.1016/0550-3213(81)90423-5}
  {\path{doi:10.1016/0550-3213(81)90423-5}}.

\bibitem{Gliozzi:2010jh}
F.~Gliozzi, M.~Pepe, U.-J. Wiese, {Linear Broadening of the Confining String in
  Yang-Mills Theory at Low Temperature}, JHEP 1101 (2011) 057.
\newblock \href {http://arxiv.org/abs/1010.1373} {\path{arXiv:1010.1373}},
  \href {http://dx.doi.org/10.1007/JHEP01(2011)057}
  {\path{doi:10.1007/JHEP01(2011)057}}.

\bibitem{Luscher:2001up}
M.~Luscher, P.~Weisz, {Locality and exponential error reduction in numerical
  lattice gauge theory}, JHEP 0109 (2001) 010.
\newblock \href {http://arxiv.org/abs/hep-lat/0108014}
  {\path{arXiv:hep-lat/0108014}}.

\bibitem{Koma:2003gi}
Y.~Koma, M.~Koma, P.~Majumdar, {Static potential, force, and flux tube profile
  in 4-D compact U(1) lattice gauge theory with the multilevel algorithm},
  Nucl.Phys. B692 (2004) 209--231.
\newblock \href {http://arxiv.org/abs/hep-lat/0311016}
  {\path{arXiv:hep-lat/0311016}}, \href
  {http://dx.doi.org/10.1016/j.nuclphysb.2004.05.024}
  {\path{doi:10.1016/j.nuclphysb.2004.05.024}}.

\bibitem{Parisi:1983hm}
G.~Parisi, R.~Petronzio, F.~Rapuano, {A Measurement of the String Tension Near
  the Continuum Limit}, Phys.Lett. B128 (1983) 418.
\newblock \href {http://dx.doi.org/10.1016/0370-2693(83)90930-9}
  {\path{doi:10.1016/0370-2693(83)90930-9}}.

\end{thebibliography}

\end{document}